# Terahertz scanless hypertemporal imaging

L. Zanotto[1], G. Balistreri[1], A. Rovere[1], O-Pil Kwon[2], R. Morandotti[1], R. Piccoli[1,3,*], L. Razzari[1,*]

[1] *Institut National de la Recherche Scientifique, Centre Énergie Matériaux Télécommunications (INRS-EMT), Varennes, Quebec J3X 1S2, Canada*

[2] *Department of Molecular Science and Technology, Ajou University, Suwon, 443–749, South Korea*

[3] *Politecnico di Milano, Department of Physics, Piazza Leonardo Da Vinci, 32, I-20133, Milano, Italy*

*[riccardo.piccoli@polimi.it](mailto:riccardo.piccoli@polimi.it) ; *[luca.razzari@inrs.ca](mailto:luca.razzari@inrs.ca)

**Since its first demonstration in 1995[1], terahertz time-domain imaging has attracted an increasingly growing interest for its ability to reveal spectral fingerprints of materials, probe changes in refractive index and absorption, as well as detect the inner structure of complex objects via time-of-flight measurements. Practically, however, its widespread use has been hampered by the very long acquisition time typically required to spatially raster-scan the object and, for each spatial point, record the field in time via a delay line. Here, we address this fundamental bottleneck by implementing a scanless single-pixel imaging scheme, which sets the path for an unprecedented reduction of both system complexity and acquisition time. By properly exploiting natural wave diffraction, time-to-space encoding applied to terahertz point detection allows for an almost instantaneous capture of the terahertz waveforms, while multidimensional images are reconstructed via a computational approach. Our scheme is a promising solution for the development of next-generation fast and compact terahertz imagers perfectly suitable for high-repetition-rate laser sources.**

During the last thirty years, terahertz (THz) technologies have experienced a steady growth and today are recognized as powerful tools for interrogating materials, since they provide information complementary to that obtained using microwaves, infrared/visible/ultraviolet (IR/VIS/UV) light, and X-rays. THz radiation has some unique features, such as the ability to pass through many optically-opaque materials (e.g. plastics, semiconductors, papers)[2], allowing for the exposure of concealed objects[3] or the study of layered structures[4], as well as a low photon energy (meV), which makes it intrinsically safe to use. In particular, time-domain spectroscopy (TDS), arguably the most powerful THz experimental technique, enables the coherent detection (i.e., the retrieval of both amplitude and phase) of the electric field associated with THz pulses[5]. As a result, it allows for the full characterization of the optical properties of an arbitrary material (e.g., the complex refractive index)[6], as well as the probing of rotational/vibrational transitions or collective excitations, such as phonons, characteristic of a variety of substances at THz frequencies (e.g., gases[7], liquids[8], organic[9]/inorganic[10] materials, bio-molecules[11]). Moreover, the picosecond duration of THz pulses enables their use to perform time-resolved measurements of certain phenomena occurring on very short time scales[12], or even to uniquely drive and control certain others[13]. The extension of these capabilities to imaging is thus a very promising technology for a large variety of purposes, including security[14], quality and safety control in industry[3,15], biomedical applications[16,17], and also art conservation[18,19].

In contrast to popular detection techniques based on photo-/thermo-electric effects, a number of novel approaches have been developed to directly retrieve the electric field associated with a THz pulse. However, this often requires sampling the THz waveform with a shorter and variably delayed optical gating pulse[20–22], which makes the acquisition process rather slow and cumbersome. Additionally, for imaging applications, the object typically has to be also raster-scanned pixel-by-pixel, thus rendering the technique complex to implement beyond research settings[23]. To address both issues and pave the way for THz hypertemporal imaging in real time, we propose a scheme that combines the single-pixel imaging (SPI) method in space, to avoid raster-scanning the object, and a scanless detection technique in time, to avoid the temporal reconstruction of the THz waveform by means of a delay line (see simplified schematic in Fig. 1).

SPI was first demonstrated in the work "Dual photography" by Sen et al.[24] in 2005, and has fostered a great deal of research ever since. SPI enables the reconstruction of an image using a detector without spatial resolution, by measuring the correlation between a set of light patterns and the spatial features of the object under investigation. Such correlation is evaluated by simply recording the transmitted (or reflected) light with the single-pixel detector. A reconstruction algorithm subsequently enables the retrieval of the original image. Afterwards, Duarte et al.[25] identified SPI as a suitable technique to implement the concept of compressed sensing (CS) for imaging purposes. CS describes the conditions under which it is possible to reconstruct a signal from an under-sampled dataset[26]. The compressibility of an image is related to the concept of "sparsity", which directly connects to the number of non-zero entries in the vector representing the image. The implementation of CS in SPI-based systems makes it possible to significantly boost the speed performance when detector arrays are unavailable or very expensive, such as in the IR or deep UV regions[27]. One such case is that of THz radiation, which is the reason why THz SPI has become a widely explored solution (ref. [28] and references therein).

It is instructive to connect the mechanism underlying SPI and the wave nature of light. The well-known Huygens-Fresnel principle elegantly catches the physics behind diffraction[29]. According to this principle, the complex field amplitude in a single point after free-space propagation can be interpreted as a summation of all the spherical waves generated at every point of the source plane (see supplementary). Therefore, even though one might believe that diffraction is a detrimental effect that befogs image reconstruction, we show here how the correct operation of our scheme strictly relies on its existence. As mentioned earlier, thanks to the structured illumination characteristic of SPI, we only need a measurement of the correlation between patterns and the object spatial shape. Diffraction automatically performs the spatial integration that brings the correlation information to a single point, thus naturally delivering the working conditions for SPI. We can see how this occurs by putting the problem in mathematical terms. The diffraction solution for a propagating patterned beam can be reduced to a spatial integral of the product between the $n^{th}$ pattern distribution, $P_n(\xi, \eta)$, and the transmission (reflection) function of the object, $T(\xi, \eta)$:

$$W_n \propto \iint_\Sigma P_n(\xi,\eta) T(\xi,\eta) d\xi d\eta \qquad (1)$$

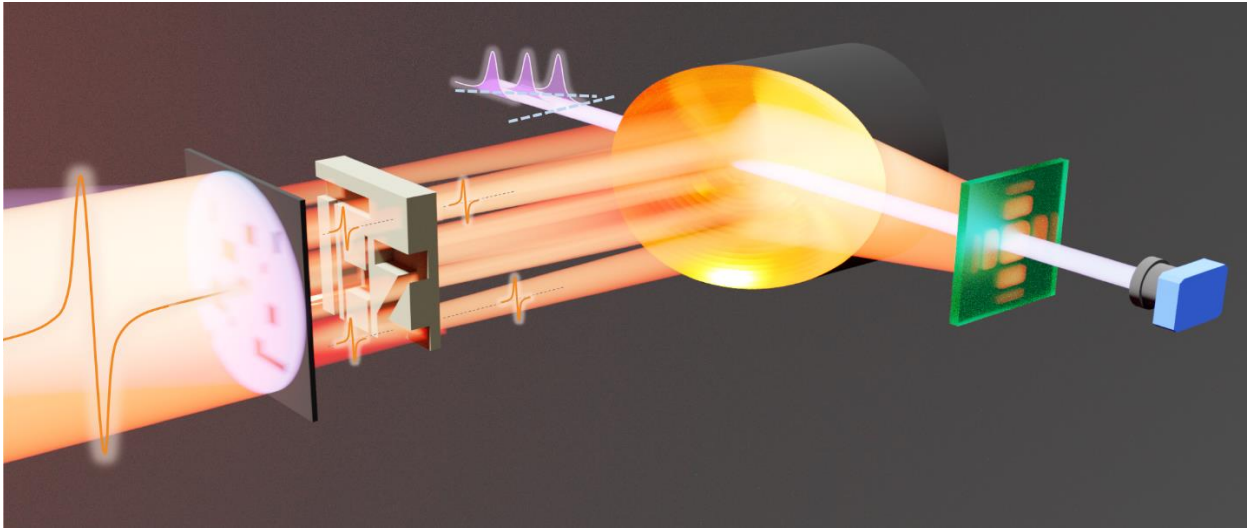

**Figure 1 | Illustration of the scanless-SPI technique.** The incoming THz pulse (in orange from left) is first spatially modulated, then propagates through the semitransparent sample ("THz" shape), and eventually reaches the detection crystal after being focused by an off-axis parabolic mirror. The probe pulse (featuring a tilted front) samples the center of the spatial Fourier transform of the THz wave distribution in the focal plane of the parabolic mirror, where the electro-optic sampling detector is placed (green crystal on the right). The probe spatial intensity is finally recorded by a camera (blue device on the right).

$W_n$ is the point value of the electric field at the center of the observation plane and corresponds to the quantity recorded by the single-pixel detector for the $n^{th}$ pattern (see supplementary for a detailed mathematical derivation). The integral in Eq. 1 gives a measure of the correlation between $P_n(\xi,\eta)$ and $T(\xi,\eta)$, being therefore mathematically equivalent to an SPI problem. In principle, then, simple free-space propagation suffices to perform an SPI measurement, yet, typically, a focusing element is employed to concentrate the energy and improve the recorded signal-to-noise ratio (SNR) (see supplementary). In addition, since at THz frequencies coherent detection directly enables the retrieval of the complex value of the electric field at the detection point, complete (i.e., amplitude and phase) image reconstruction can be achieved, even in the most general case where all the entities involved (i.e., patterns and object transmission/reflection) are complex-valued. The phenomenon of diffraction has been exploited in various SPI-based arrangements operating in the VIS/IR range, like in systems for lens-free diffractive imaging[30], as well as in setups for complex field retrieval utilizing ptychography[31] and digital holography[32].

In our imaging system, a scanless THz detection technique (also typically referred to as "single-shot", for its ability to potentially operate with an individual pulse) has been judiciously coupled to SPI by means of diffraction, to address the issue of the slow and complex point-by-point temporal scan needed to obtain the THz waveform. Indeed, while an SPI-based THz imaging system has recently reached real-time operation for fixed time amplitude measurements[33], the full potential of THz time-domain imaging in practical implementations is still untapped, due to the limits imposed by the temporal scan. Traditionally, this scan is performed via a mechanical delay line, which varies the delay of a probe pulse relative to the THz wave. Such delay lines are based on mechanical moving parts that are prone to failure and significantly slow down the acquisition process. All-optical[34] and acousto-optical[35] delay lines with sampling rates up to hundreds of Hz and tens of kHz, respectively, have been recently developed. These systems enable fast operation without moving parts, yet they require the use of two accurately synchronized lasers in the first case, or a dedicated acousto-optical modulator and sophisticated electronics in the second, which considerably increase complexity and cost of the imaging system. During the last twenty years, single-shot techniques have been explored as tools to simultaneously record the whole THz waveform in the time-domain. They are generally based on stretching the probe pulse temporally or spreading it spatially, so to enable the simultaneous probing of different temporal positions of the waveform[36]. For our implementation, once again diffraction represents a key aspect. To achieve a "true" "single-point" detection[37] and satisfy the conditions under which Eq. 1 is valid, the probe size at the detector plane should be smaller than the width $w$ of the main lobe of the THz diffracted pattern at the same plane[29]: $w = \lambda_{THz} z / x_{MAX}$ ($w \approx 1.5$ mm in our specific case), where $\lambda_{THz}$ is the THz wavelength, $z$ is the propagation distance, and $x_{MAX}$ is the scene maximum spatial extent. This is reasonably simple to achieve using a probe beam at IR/VIS wavelengths, about 3 orders of magnitude shorter than that of THz radiation. By keeping the probe smaller than this size, in a somehow counter-intuitive fashion, we can sacrifice one spatial dimension to allocate the time domain (or, in other words, use a time-to-space encoding technique[38]), yet still obtaining a complete spatial reconstruction of the object to be imaged, while avoiding the strict limitations in temporal resolution that would be imposed by a time-to-frequency encoding strategy[39]. Time-to-space mapping is based on tilting the probe pulse front so that spatially separated points arrive at the detector plane at different times, therefore simultaneously probing different temporal positions of the THz pulse. Thereafter, the probe spatial intensity is captured with a camera, thus obtaining an in principle instantaneous picture of the THz temporal waveform.

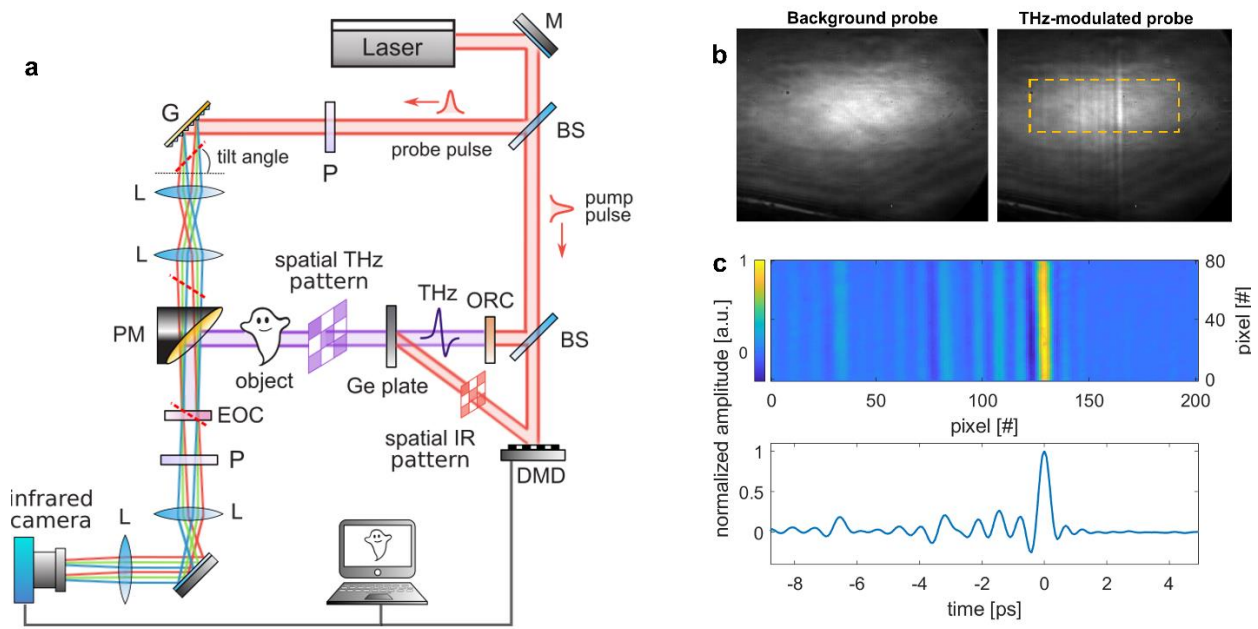

**Figure 2 | Operation of the THz scanless hyperimaging system.** (a) Experimental setup. The laser beam is split into two lines (pump and probe) using a beam splitter (BS). The pump pulse is then split again: one line generates THz radiation via optical rectification in an organic crystal (ORC), the other is spatially modulated using a digital micromirror device; the patterned infrared beam is used to photo-excite carriers in a Ge window, in order to spatially modulate the THz pulse; the latter interrogates the sample and then reaches the detection crystal (EOC). The probe line goes through a polarizer (P) and then hits a grating (G), which tilts the probe pulse front; two lenses (L) form a 4-f system that images the tilted probe beam onto the detection crystal (EOC). After such crystal, an analyzer and a second 4-f system take the THz-modulated probe beam to the infrared camera. (b) Examples of images recorded by the infrared camera: background probe beam (left); THz-modulated probe beam (right). The dotted region corresponds to the area in (c). (c) (top) Portion of the area extracted from the images recorded by the camera and corrected by subtracting and dividing the modulated probe by the background; (bottom) THz waveform reconstructed by integrating the image above along the vertical direction.

Our experimental setup is illustrated in Fig. 2(a). A third line is added to the two employed for THz generation and probing, where a fraction of the laser beam is spatially patterned by means of a Digital Micromirror Device (DMD - see methods) and then used for THz photo-modulation[40]. In the probe beam line, a diffraction grating is used to tilt the probe pulse front. Such tilted probe is then imaged onto the electro-optic detection crystal to enable the capture of the THz waveform in the time-domain, and finally propagates to reach the IR camera (see methods). This eliminates the need for the temporal scan typically performed with a delay line. In Fig. 2(b), we can see examples of probe images recorded by the IR camera. The waveform is extracted by subtracting the background probe image from the one modulated by the THz beam. A proper normalization is also performed, to correct for the probe intensity profile (see methods). In Fig. 2(c), we show an example of the retrieved image used for the reconstruction, as well as the actual THz waveform obtained by integrating along the vertical direction to improve the SNR.

THz image reconstruction is achieved thanks to our time-domain SPI algorithm, implemented by solving an SPI problem at every point of the time window, as explained in detail in ref.[41]. In this way, the transmitted THz waveform corresponding to each image "pixel" can be obtained, thus retrieving the exact same information that would be acquired with a traditional raster-scan-based TDS imaging system. To prove the potential of the proposed imaging scheme, in what follows we show the reconstruction of 3D images of multilayered samples, by displaying the layer contrast with different methods.

We employed two samples made of high-density polyethylene (HDPE), which has a refractive index of 1.54 and negligible absorption ($\alpha \leq 1$ cm$^{-1}$) in the range 0.5-2.5 THz[42]. Sample (1) (Fig. 3(a)) has the letters T-H-Z engraved with three different thicknesses (T – 1 mm, H and Z – 2 mm, outside area – 3 mm), while sample (2) (Fig. 3(b)) the letters I-N-R-S (I and N – in air, R and S – 1 mm, outside area – 2 mm).

We first reconstructed images while considering the THz field amplitude at various fixed time positions. In Figs. 3(c),3(d) we show three $32 \times 32$ frames, retrieved by taking, at every pixel, the normalized field amplitude at the time positions of the three peaks in Figs. 3(e),3(f). Each image shows the spatial features of the layer with a thickness corresponding to that time delay (see also the full time-domain video in supplementary material). The in-plane resolution is dictated by the pixel size utilized to generate the modulation patterns (312.5 µm in this case). We can see that the reconstruction reproduces well the object shape, excluding some portions at the letter and sample edges, as well as some smaller features, where probably scattering effects play a role in reducing transmission.

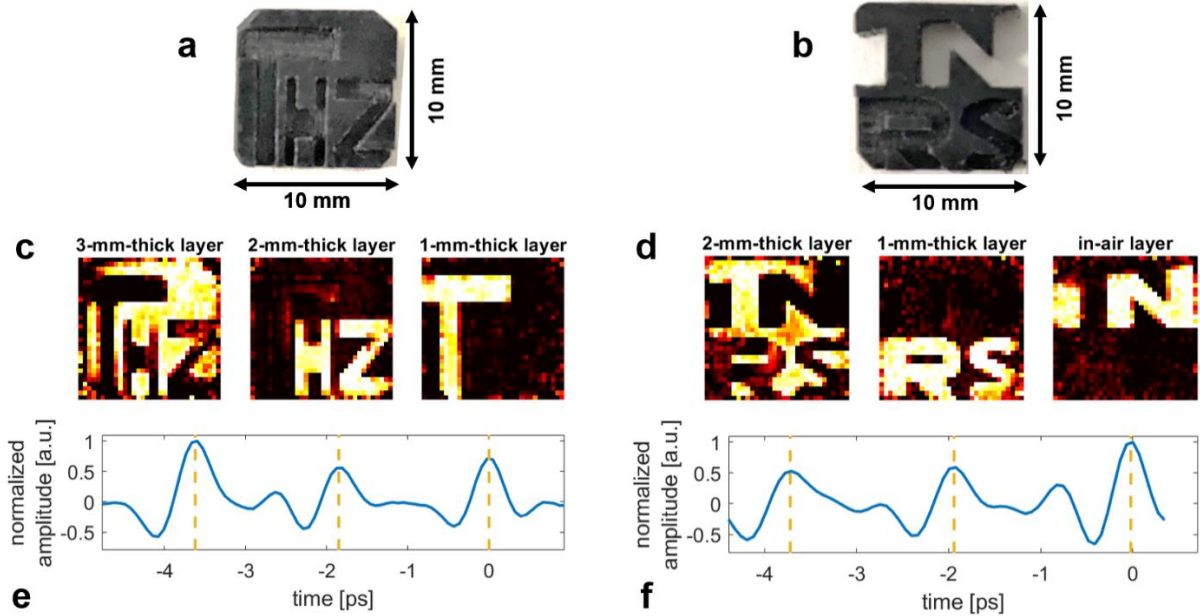

**Figure 3 | Image reconstruction.** (a) sample (1), high-density polyethylene (HDPE) with T-H-Z letters engraved. (b) sample (2), HDPE with I-N-R-S letters engraved. (c)-(d) Images reconstructed considering the electric field amplitude at the time positions corresponding to the three different peaks of the THz waveforms in (e)-(f) respectively. The waveforms in (e)-(f) are examples of measured THz traces before the SPI reconstruction is applied. The multipulse nature of these waveforms is due to the combined contribution from all the "white" pixels of the utilized pattern.

For the same samples, we also measured the time delay between a reference THz pulse propagating in air and the THz pulses retrieved at each pixel. We therefore obtained images where the pixel colour displays the time delay experienced by the THz pulse at that position (incidentally, by considering the HDPE group refractive index, it is also possible to directly extract the relative thicknesses of the different layers). In Fig. 4, we present both $16 \times 16$ and $32 \times 32$ images for the two samples. The time-of-flight reconstruction of 16 x 16 images is accurate and allows to discern the letters at each layer, although the limited resolution (625 µm) hinders the precise reconstruction of the smallest features. In the 32 x 32 images, the increased resolution clearly enables to better show the sharp spatial features. However, we also note that some pixels, mostly at the edges of the letters and the sample, fail to represent the proper time delay, most likely due to scattering as well as diffraction effects. Indeed, the pixel width is in this case 312.5 µm, thus comparable to the central wavelength (300 µm at 1 THz) of the THz pulses. This means that even after very short distances, the propagating beam will undergo diffraction modifications (Rayleigh range ≈ 0.3 mm at 1 THz), which alter the pattern spatial distribution interrogating the sample, in turn leading to a potential reduction of the resolution of the reconstructed image. Even though this effect could be numerically corrected by means of post-processing techniques[43], here we have limited it by using a thin Ge wafer (125 µm) and by placing the sample directly after it.

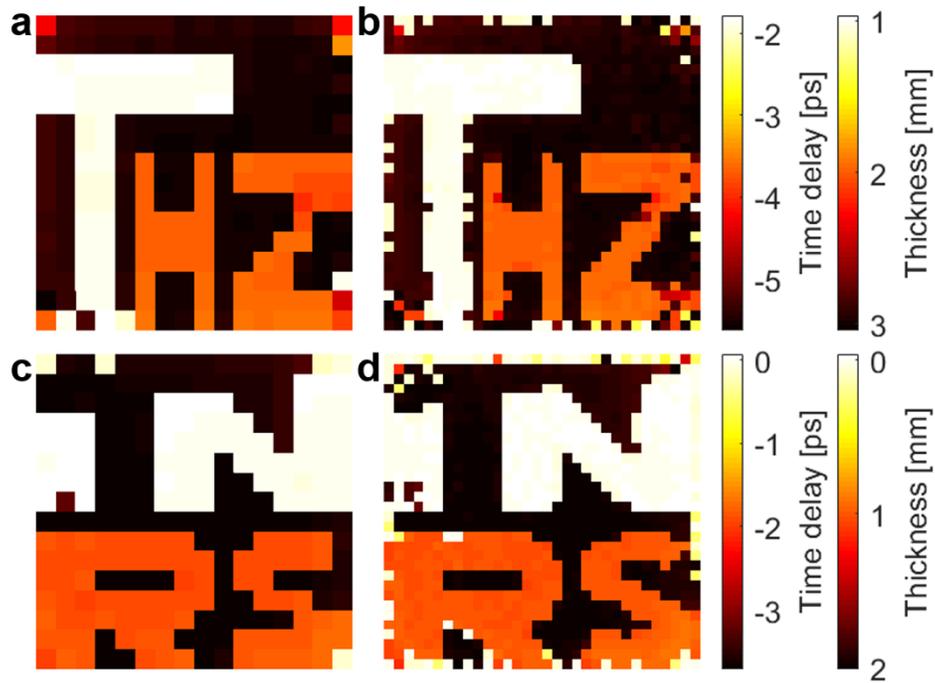

**Figure 4 | Time-delay images.** (Top) Images of sample (1): (a) 16 × 16 and (b) 32 × 32 pixels images reconstructed considering the time delay between the THz pulse peak in air and the peak of the transmitted pulses retrieved at each pixel position. (Bottom) Images of sample (2): (a) 16 × 16 and (b) 32 × 32 pixels images reconstructed in the same way. The two colors bars represent the temporal delay (left) and the local thickness of the samples (right).

As a proof-of-principle, we also implemented a standard non-iterative CS algorithm[44] to reconstruct the image using a number of patterns that is smaller than the total number of pixels (1024 in the case considered here). As shown in Fig. 5, even using only 25% of the patterns (compression ratio, CR = number of patterns used/number of pixels), it is still possible to retrieve a good approximation of the spatial features of the three layers. Better compressions could be in principle achieved by using different pattern sets[45] or more advanced reconstruction algorithms[44]. These solutions have been proved to be effective at optical frequencies but are not always straightforward to implement in the THz range, due to the limitations posed by the photo-modulation technique (for instance, the complexity in generating non-binary patterns). Innovative methods for THz modulation, potentially more flexible, have been introduced in recent years[46,47], but still need further development to become viable solutions.

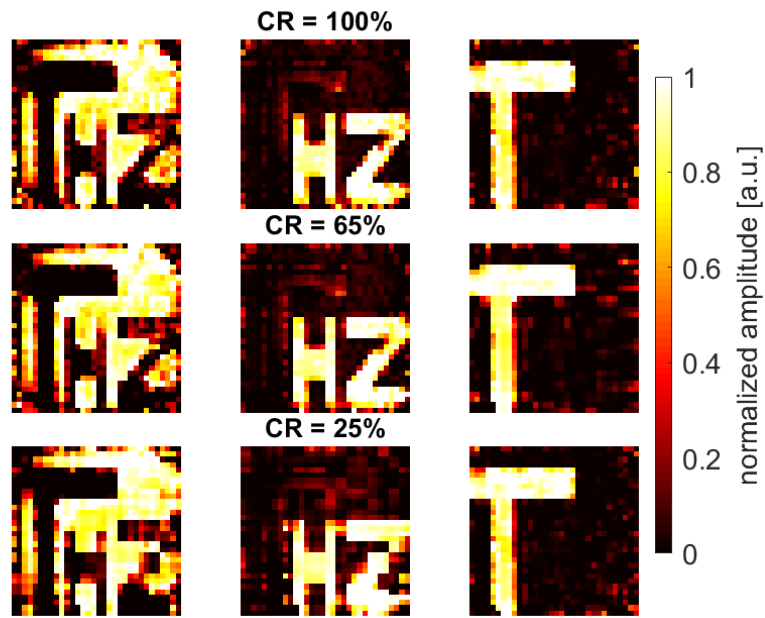

**Figure 5 | Compressed sensing reconstruction.** 32 × 32 images of sample (1) reconstructed considering the amplitude at different time positions (as in fig. (3)), and with a decreasing CR (from top to bottom, 100 % - 65 % - 25 %).

It is important to underline that, in the imaging examples shown above, from the same time-domain data we alternatively extracted the local transmitted amplitude as well as the delay relative to a reference pulse, thus making use of both amplitude and phase information. This is a demonstration that our imaging implementation fully preserves the capabilities of TDS coherent detection: depending on the type of sample under investigation, different contrast methods can thus be selected, so combining spatial 2D reconstruction with the extraction of, e.g., the complex refractive index, time-of-flight information or even spectral absorption lines[41]. At the same time, our arrangement is cost-effective and extremely simple, basically requiring, besides the laser source, just a DMD and a camera to be operated, without the presence of any complex device (mechanical or not) else needed to perform the spatial/temporal scans.

The acquisition of a 32 × 32 image (1024 patterns) using our current system still requires a long time (~40 min.), mainly due to the need of a long average for each frame captured by the camera (2.5 s) (we note that the acquisition of the same image without the single-shot technique would take ~40 hours via our setup and a traditional delay line). Such time-consuming acquisition is necessary to obtain a sufficient SNR at the limited repetition rate of our THz source (250 Hz). Using laser systems featuring repetition rates in the MHz range, coupled with photoconductive antennas for THz generation, the same SNR could be obtained by capturing ms-long frames, potentially bringing the acquisition time of the whole hypertemporal image down to few seconds. Moreover, as shown above, the application of CS algorithms could further reduce such time, by limiting the number of patterns to a small fraction of the full set, towards real-time operation.

In conclusion, we have presented a novel configuration for THz multidimensional imaging that combines the SPI scheme with a scanless time-domain detection technique, so to achieve fast imaging without the need for complex electronics and any bulky mechanical moving part. From a more fundamental viewpoint, we have shown that diffraction plays a crucial role in the proposed technique. In particular, the scanless single-point THz coherent detection implemented in this work counter-intuitively exploits time-to-space mapping at the detection plane to fully retrieve multidimensional images (time/frequency, in addition to space), with the further possibility of extending the technique to complex-valued patterns and objects without significant limitations. We envision that coupling our scheme to high-repetition rate laser sources, such as those based on ytterbium-doped media, could lead to compact THz hypertemporal imagers operating in real time.

# Methods

## *Hadamard pattern set and "cake-cutting" ordering*

The choice of the set of patterns typically depends on the operating conditions (e.g. light source wavelength, resolution, source brightness). One of the most used pattern bases is the Hadamard set, which guarantees an optimal SNR[48]. The Hadamard set is orthonormal and can be obtained with an iterative procedure[49]. In the present work, its construction was achieved by means of a built-in Matlab® function. To implement a fast compressed sensing reconstruction, we used the Hadamard set ordered in a "cake-cutting" sequence[49]. This ordering ensures that the most significant patterns are always displayed first, leading to a better reconstruction with lower sampling ratios and at a lower computational cost.

## *Time-to-space encoding and near-zero transmission detection*

The time-to-space scanless technique employed in this work is based on tilting the probe pulse front in such a way that spatially separated points arrive at the detection crystal at different times, therefore probing different temporal positions of the THz waveform simultaneously. The pulse front tilt is achieved by means of a dispersive element (e.g., transmission/reflection grating), which spatially separates different frequency components of the broadband probe pulse, and a 4-*f* system enabling the formation of an image of the (tilted) probe front at the detection crystal position. The resulting time window, $T$, is a function of the probe beam diameter, $\sigma$, at the detection position, and the tilt angle, $\gamma$. The tilt angle can, in turn, be written in terms of the mean wavelength of the light used, $\bar{\lambda}$, the grating parameter, $d$, the magnification of the 4-*f* system, $M$, and the diffraction angle after the grating, $\beta$ ($c$ is the speed of light in vacuum)[50]:

$$T = \frac{\sigma \tan \gamma}{c} = \frac{\overline{\lambda}\,\sigma}{c\,M\,d\,\cos\beta} \qquad (1)$$

In the next paragraph, quantitative estimations for our experimental conditions are presented.

The detection of the THz pulses is performed in a near-zero transmission configuration[51]. In more detail, a linear polarizer is placed after the detection crystal with its axis almost perpendicular to the input probe polarization direction, leading to a probe transmission close to zero. When the THz pulse is present, an electric-field-induced birefringence in the crystal (Pockels effect) changes the probe polarization state, thus modifying the intensity transmitted through the polarizer. In the single-shot setup, different points along the probe pulse front undergo a different polarization modification, corresponding to a different temporal position along the THz waveform. The probe beam is then imaged onto the camera and its intensity variations reveal the THz waveform temporal shape. The waveform is extracted by first subtracting the image of the background probe from the one modulated by the THz, and then dividing the result by the background, to correct for the gaussian shape of the probe beam:

$$\Gamma = \frac{I_{THz} - I_{bg}}{I_{bg}} \qquad (2)$$

*Experimental setup*

The output from an amplified Yb laser (1030 nm, 175 fs pulse duration, 1 mJ at 250 Hz repetition rate) is split into a pump beam, a probe beam, and a pattern generator by means of a 4:96 wedge window and 50:50 beam splitter. The pump beam is used to produce THz pulses via optical rectification in a 300-μm-thick HMQ-TMS (2-(4-hydroxy-3-methoxystyryl)-1-methylquinolinium 2,4,6-trimethylbenzenesulfonate) organic crystal. After some magnifying optical elements, the THz beam propagates through a 125-μm-thick germanium plate (undoped, resistivity >40 Ωcm) used for spatial modulation, and eventually passes through the sample (with a 10 mm × 10 mm size). Finally, free-space electro-optic sampling is carried out via a 3-mm-thick gallium phosphide (GaP) crystal, where the THz and probe beams are spatially overlapped. The SPI scheme requires spatial patterns to interrogate the sample. To obtain such patterns on the THz beam, we applied a modulation technique based on carriers photo-excitation in a semiconductor, first described in[40]. The required IR pattern generation beam is modulated by means of a DMD (LightCrafter4500, Texas Instruments) and illuminates the surface of the Ge plate, locally exciting photo-carriers. The THz pulse passing through the wafer shortly after the excitation is screened where carriers are present, but transmitted (albeit partially reflected) elsewhere. By applying this technique, we achieve a modulation depth of about 90% (peak amplitude of the THz pulse). To implement the scanless (single-shot) detection scheme described above, a reflection grating (1200 grooves/mm) is placed along the probe beam path, followed by a 4-$f$ system (50 cm and 10 cm focal lengths), which images the probe onto the detection crystal with a reduced size (1 mm diameter) and a tilted front. After sampling the THz pulse, the probe is magnified by a second 4-$f$ system (5 cm and 50 cm focal lengths) and then reaches the recording InGaAs camera (Xeva 320 Series, Xenics). With the grating employed and operating at an incidence angle of 72° (diffracted angle 16°), the expected available time window provided by the probe is ~19 ps, which is suitable to properly acquire the THz waveforms replicas transmitted through the multilayered samples. We time calibrated the scanless detection by means of a mechanical delay line, by changing the probe optical path and checking the corresponding shift of the spatial position of the THz pulse peak on the camera. The temporal resolution obtained in our experimental conditions was 68.7 fs/pixel.